\documentstyle[aps,preprint,epsfig]{revtex}
\tighten

\csname @twocolumnfalse\endcsname

\begin{document}
\newcommand{\bea}{\begin{eqnarray}}
\newcommand{\beq}{\begin{equation}}
\newcommand{\eea}{\end{eqnarray}}
\newcommand{\eeq}{\end{equation}}

\title
{Conductance of Open Quantum  Billiards and
Classical Trajectories}

\author
{R.G. Nazmitdinov$^{1,2}$, K.N.~Pichugin$^{3,4}$,
I. Rotter$^1$ and
P.~${\rm \check S}$eba$^{3,5}$}

\address{
$^1$ Max-Planck-Institut f\"ur Physik komplexer
Systeme, D-01187 Dresden, Germany\\
$^2$ Joint Institute for Nuclear Research, 141980 Dubna, Russia\\
$^3$ Institute of Physics, Czech Academy of Sciences,
 16253 Prague, Czech Republic \\
$^4$ Kirensky Institute of Physics, 660036 Krasnoyarsk, Russia\\
$^5$ Department of Physics, University Hradec Kralove, 
50003 Hradec Kralove, Czech Republic
}               
\date{\today}
\maketitle

\begin{abstract}
We analyse the transport phenomena  of  2D  quantum
billiards with convex boundary of different shape. The quantum mechanical
analysis is performed by means of the poles of the $S$ matrix while the
classical analysis is based on  the motion of a 
free particle inside the cavity along trajectories with a different
number of bounces at the boundary. The value of
the conductance depends on the manner the leads are attached to the cavity.
The Fourier transform  of the transmission amplitudes
is compared with the length of the classical paths. 
There is good agreement between classical and quantum mechanical results 
when the conductance is achieved mainly by special
short-lived states such as
whispering gallery modes (WGM) and bouncing ball modes
(BBM). In these cases, also the localization of the wave functions
agrees  with the picture of the 
classical paths. The $S$ matrix is calculated classically
and compared with the
transmission coefficients of the quantum mechanical calculations for five
modes in each lead. The number of modes coupled  to the special
states is effectively reduced.
\\
\end{abstract}


\section{Introduction}

The problem whether and how classical dynamics of mesoscopic systems 
is manifest in quantum mechanical characteristics is 
studied intensively during the past decade.
It is well established that the statistical fluctuations of  
quantum systems whose associated classical dynamics is chaotic   
are well described by random matrix theory,  see e.g.
the recent reviews \cite{ben,alh}.
This approach treats the spectra  of many dense lying  states 
by means of statistical methods neglecting
the individual properties of the states \cite{mello}.
 
In quantum systems with low level density, however, 
deviations from the randomness are observed and discussed
\cite{ro91,tar,ralph,alt,a1,a2,berg,burg,pra,berg2}.
The results  point at   quantum mechanical interference effects
between the  quantum states, which may
become important under certain conditions.
These effects are displayed, e.g., in the transport phenomena 
through  quantum dots,
when the leads are configurated in such a manner that
one or few propagating modes are supported \cite{a1,a2,berg}.
The underlying processes are not fully understood, up to now.
A detailed analysis of the internal structure 
of the corresponding Hamiltonian is therefore required. 
Here,  new questions arise such as (i)
which role play the individual properties of the states whose small
number in a certain energy region does, generally, not allow a 
statistical description, and (ii)
which properties of the states survive when the system is 
embedded into an environment. 

While at weak
coupling between system and environment the last question 
is believed to be answered, it is still open at strong coupling
where  the influence of the channels onto  the states may be  large.
A study of these problems in real systems is difficult since 
their separation  from other questions such as many-body correlations  
and the shape of the effective potential is impossible. The most 
transparent answers are expected from a study of microwave cavities 
which simulate well the features of real quantum systems \cite{stock}. 
In this case, the shape of the system
is well defined and two-body forces do not exist.

Theoretical and experimental studies on open microwave cavities 
and also on quantum dots at low energy 
have shown that  the individual properties of the states and their
matching to the wave functions of the environment play an important role,
indeed \cite{a1,a2,pese,ropepise,1,a4}. Analytical considerations show 
that level repulsion as well as  level clustering may appear. 
The repulsion of the states in
energy is accompanied by  adjusting their widths 
(inverse lifetimes of the states) while the clustering of levels
is accompanied by a bifurcation of the widths.
Both phenomena are observed, in fact, in numerical studies on rectangular
billiards in which the  matching of the wave functions
is varied  by means of enlarging (or reducing) the area of 
the cavity \cite{ropepise}. 
Clusters in the tunnelling resonance spectra of 
ultra-small metallic particles of a few nanometer size have been
observed experimentally \cite{ralph} and explained theoretically \cite{alt}. 

The wave-function statistics for ballistic quantum transport through chaotic
open billiards is investigated in \cite{berg2}. Here,  the 
chaotic-scattering wave
functions in open systems are interpreted quantitatively in
terms of statistically independent real {\it and} imaginary random fields 
in the same manner as for wave-function statistics of closed 
systems. This result may be
compared with a similar one obtained from an analysis of 
the nuclear coupling to the one-channel continuum \cite{drokplro}. 
The Gaussian distribution of both, the real and imaginary parts, 
seems therefore to be a 
common feature of the wave-function statistics of small open quantum systems.
Such a result does not follow from the random matrix theory.

The role of the matching of the wave functions for the dynamics of
the system is studied further in \cite{1}. Here, some special states
are shown to accumulate the total coupling strength between system and
environment which is expressed by the sum of the widths
of all states lying in the energy region considered. The accumulation takes
place by resonance trapping, i.e. all states but the special ones decouple
more or less from the environment while the widths of the special states
reach the maximum possible value. 

The quantum billiard considered in \cite{1} has the shape of a semicircle with
an internal scatterer (SIS). Here, bands of overlapping resonance
states appear
whose wave functions are localized either along the convex boundary 
of the cavity or along
the direct connection between the two attached leads. The first type of
resonances is related to whispering gallery modes (WGM) and the other one to
bouncing ball modes (BBM). The transition from one type to the other is traced
in \cite{1} by varying the position of one of the two attached leads. As a
result, the BBM being special states at a certain position of the attached 
leads, are trapped by the WGM at another position of the leads.
The internal scatterer in the SIS does not play any role for this phenomenon
since it appears in a quantum billiard of semicircle shape without any
internal scatterer as well.
Meanwhile, the phenomenon of resonance trapping has been proven experimentally
\cite{perostba}. 

Thus, some
special states survive at strong coupling of the system to the
environment and give, under certain conditions, a large contribution to 
observable values, e.g. to the transmission (conductance).
Besides these special states there exists, at the same energy, a large 
number of long-lived states
that are decoupled more or less from the environment and  
contribute incoherently to the observables. 
The coherent contribution to the
conductance is considered mostly as background conductance and the
incoherent contributions create fluctuations on this background.  
The transmission shows  a gross structure (background) caused by the 
special states
and a fine structure (fluctuations) created by the interferences with the 
long-lived trapped states. Accordingly, a 
Fourier analysis of the transmission spectrum 
contains information not only on the long-lived states but also on the special
states in spite of their background character. 

In the present paper, 
we consider  quantum billiards of Bunimovich type with
different positions of the attached leads. Since the
closed Bunimovich billiard shows the features of chaotic dynamics,
this system is especially suited for the study of the question which 
states survive after embedding it into an environment. We will show 
that an appropriate attachment of the leads selects special states 
which enhance  the conductance as compared to the predictions of 
random matrix theory. Further, 
we  compare the results of a Fourier analysis of the transmission spectra 
with the results of classical calculations for the conductance of 
cavities having the same geometry. 
This comparison  will provide us  information on the 
question to which degree  the classical 
properties of dynamical systems are manifest in  quantum mechanical 
characteristics, in particular in  the phenomenon of 
transport through billiards
with both a small number of states and a small number of open channels.

The paper is organized as follows.
In Sect. II, the basic equations underlying the quantum mechanical description
are given. In Sect. III, we provide the results obtained numerically for 
quantum billiards of Bunimovich type  to which the 
leads are coupled in a different manner. They are configured to support a 
small number of propagating modes ($Z\leq 5$).
We represent the eigenvalue pictures together with some wave functions and the
power spectra obtained from the Fourier analysis of the transmission and
reflection fluctuations. The values are compared with those calculated 
classically. Furthermore, the  $S$ matrix is calculated classically 
and compared with the
transmission coefficients of the quantum mechanical calculation for five
modes in each channel.
The results are discussed in Sect. IV and summarized in the last section.

\section{Basic equations of the quantum mechanical description}
\label{basic}

We consider a two-dimensional flat resonanator
coupled to two leads and solve the 2D Schr\"odinger equation
\beq
\label{sr}
-\frac{\hbar^2}{2 m}\Delta \Psi=E\Psi
\eeq
under the assumption that the potential is zero inside the billiard and inside
the leads but infinite outside these regions. The walls are
assumed to be  infinitely hard. In other words,
we use the Dirichlet boundary condition $\Psi=0$ on the boundary of the
billiard and of the leads. The wave functions inside the leads are
given as a superposition of plane waves,
\bea
\label{wf}
\Psi_1(x,y)=\sum_{m=1}^Z(a_{m} e^{ik_{m} x} + 
b_{m} e^{-ik_{m} x})u_{m}(y)\nonumber\\
\Psi_2(x,y)=\sum_{n=1}^Z(a_{n} e^{ik_{n}x}+
b_{n} e^{-ik_{n}x})u_{n}(y)
\eea
where we denote the two leads by 1 and 2, 
respectively, $u_j(y)= \sqrt{2/d}\; sin\Big(\frac{\pi j}{d}y\Big), \; j=n,m$. 
Further, $d$ is the width of  each lead 
and $ m (n) =1,2,...,Z$ is the number of transversal modes in  lead 1 (2). 
The wave number is  $k_n=\sqrt{2m_{\rm eff}/ \hbar^2~(E-E_n)}$ where
$E_n=\hbar^2n^2\pi^2 / (2m_{\rm eff} d^2)$  is the energy associated with the
transverse motion. At the energy $E$ the modes $n$ with $E-E_n>0$
are propagating while those with $E<E_n$  are evanescent waves.
In the following, we use the units $\hbar^2/(2 m_{\rm eff})  = 1$
and choose $d=1$.

By definition, the $S$-matrix maps the amplitudes of incoming waves
to those of the outgoing ones,
\bea
\label{s1}
b \; = \; S \; a \; .
\eea
The $S$-matrix can be written as 
(for details see \cite{ro91,ro81} and Sect II.B of Ref. \cite{ro01})
\bea
S_{cc'} = S_{cc'}^{(1)} - S_{cc'}^{(2)}
\label{s11}
\eea
where $S_{cc'}^{(1)}$ corresponds to the smooth direct reaction part and 
\bea
\label{smat}
 S_{cc'}^{(2)} = 2 i \pi \sum_{R=1}^N \frac{\tilde W_R^{c'}
\tilde W_R^c}{E - {\tilde E}_R + \frac{i}{2} {\tilde \Gamma}_R}
\quad ; \qquad \tilde \Gamma_R = 2 \pi \sum_c (\tilde W_R^c)^2 
\eea
is the resonance reaction part in pole representation. 
Here, the $\tilde W_R^c$ are the
coupling matrix elements between the wave functions
$\tilde\Phi_R$ of the  {\it resonance} states and the channel wave
functions   in the leads (where the Lippmann-Schwinger type relation 
between the wave functions of the resonance states and the eigenfunctions of
$H_{\rm eff}$ is used). The $c$ denote the channels
$m=1, ... ,Z, \; n=1, ... ,Z$. The poles of the $S$
matrix are related to the energies $\tilde E_R$ and widths $\tilde
\Gamma_R$ of the  resonance states of the billiard. 
They are lying at $E_R=\tilde E_R(E=E_R)$ (solutions of the fixed-point
equations). This
relation holds not only for  isolated resonance states but also
for overlapping ones as shown in Refs. \cite{ro91,ro81,ro01}. A similar 
representation of the $S$ matrix has been given in Ref. \cite{sok}. 
Although  $S_{cc'}^{(2)}$, Eq.  (\ref{smat}), has formally the standard form, 
it contains all the reordering processes taking place in the system when the
resonance states overlap, including
the influence of the channel channel coupling.  All these effects are 
expressed by the bi-orthogonality of the resonance wave functions and are
involved in the $\tilde W_R^c, \; {\tilde E}_R $ and   ${\tilde \Gamma}_R$.

For {\it isolated} resonances the widths of the 
states are much smaller than the distance between them. In such a case,
$\tilde E_R \approx E_R^{\rm d}$,   $~\tilde W_R^c \approx
W_R^{c({\rm d})}$, and the channels are not coupled. That means, 
the  $S$ matrix poles can be calculated with the help
of the coupling matrix elements $W_R^{c \rm (d)}$ (overlap integrals
between the wave functions $\Phi_R^{\rm d}$ of the
discrete states and the channel wave functions $u_n$ in the
leads),  with the energies 
${E}_R^d $  of the discrete states of the (closed) billiard and 
${\Gamma}_R^d = 2 \pi \sum_c (W_R^{c \rm (d)})^2$. This approximation
is justified for 
the description of $S$ matrix poles lying near  the real axis \cite{mawei}.

For {\it overlapping} resonances (i.e. when the widths 
exceed the energetical distance  between the resonances),
the $E_R=\tilde E_R(E=E_R)$ and $W_R^c=\tilde W_R^c(E=E_R)$ 
may differ strongly from the $E_R^{\rm d}$ and $W_R^{c({\rm d})}$, 
respectively, due to reordering processes
taking place in the billiard under the influence of the coupling
to the leads. For numerical examples   see Refs. \cite{pese,ropepise,1}.
Due to the reordering processes the $S$ matrix poles can not be approximated 
by using the $E_R^{\rm d}$ and $W_R^{c({\rm d})}$  in  the pole representation
$S_{cc'}^{(2)}$ as shown in a numerical study  \cite{schanz}.
Instead, the poles of  the $S$ matrix are given by   (\ref{smat}) where
the interaction of the resonance states via the continuum is 
taken into account by  diagonalizing the effective Hamiltonian
in the subspace of discrete states embedded in the continuum.
Its eigenvalues are the energies $\tilde E_R$ and widths $\tilde \Gamma_R$ of
the resonance states which determine the poles of the $S$ matrix, and the
$\tilde W_R^c$ are complex, generally.
For details see Refs. \cite{ro91,ro81,ro01}. Moreover, in Ref. \cite{stm}, 
the effective Hamiltonian for an open quantum billiard
is derived. Diagonalizing this effective Hamiltonian, numerical 
studies are performed for quantum billiards with isolated and overlapping 
resonances. The results  are in good agreement  with
experimental data obtained from microwave resonators of the same
shape \cite{stm}. In particular, the phenomenon of resonance trapping can 
clearly be seen in both the theoretical and experimental results.

Reordering processes may take place in open quantum systems  
{\it not only} between the states of  the
system which cause the  wave functions $\tilde \Phi_R$ of the resonance
states to be different from the wave functions $\Phi_R^{\rm d}$
of the discrete states. 
The strong coupling of some resonance states to the 
channel wave functions may cause also changes in the channel wave functions 
themselves because  they are  coupled   via the
resonance states. This coupling of the channel wave functions via the 
resonance states (channel channel coupling) is in complete analogy to the
coupling of the resonance states via the
channels. Both are caused by the same coupling matrix elements between the 
resonance wave functions and the channel wave functions. For details
see \cite{ro91}.  
It may occur that   wave functions of different channels 
couple so strongly that they appear effectively as a one-channel wave
function and exist together with less coupled channel wave functions.  
Thus, the  number of relevant channels may be  effectively reduced
at strong coupling between system and environment. For numerical examples on
quantum billiards see \cite{1} and for nuclei see \cite{drokplro}. 

Since the sum of the diagonal matrix elements of a matrix is
equal to the sum of the eigenvalues, we get \cite{ro91,ro01}
\bea
\label{gasu}
\sum_R \tilde \Gamma_R = 2 \pi \sum_{Rc}
(\tilde W_R^c)^2 = 2 \pi \sum_{Rc} (W_R^{c({\rm d})})^2 = \sum_R
\Gamma_R^{\rm d} 
\eea
where the $\Gamma_R^{\rm d} $ characterize the coupling of the states $R$
to the environment without taking into account 
any mixing (via the continuum) with the other states of the system.
Eq. (\ref{gasu}) gives the total coupling strength between system and
environment. 
It is basic for all redistribution processes taking
place in the system under the influence of the coupling to the
environment. This is confirmed in particular for redistributions which
happen in the quantum billiard  when  the position of the
attached  leads to the billiard is varied \cite{1}.
In this case,
\bea
\label{gasub}
\sum_R \tilde \Gamma_R = 2 \pi \sum_{Rc}
(\tilde W_R^c)^2 \approx const
\eea
since the $W_R^{c({\rm d})}$ are determined by an
integral over the region of attachment  \cite{schanz,stm} and 
remain almost unchanged by
varying the position of the attachment (if the number of states
in the cavity is not too small). It may happen that,
under certain conditions, 
\bea 
\label{tra} 
\sum_{R=1}^K \tilde \Gamma_R \approx \sum_{R=1}^M \tilde \Gamma_R 
\quad {\rm and}
\quad \sum_{R=K+1}^M \tilde \Gamma_R \approx 0 \; . 
\eea 
In such a case, the whole coupling strength is concentrated on $K<M$
special states while $M-K$ states are almost decoupled  from the
environment. This phenomenon, called {\it resonance trapping}
\cite{ro91}, is crucial for the conductance of
quantum billiards with WGM \cite{1}. The value of  $K$
may or may not be  related to the number $Z$ of open channels \cite{ro91}.
For  the WGM,  $K$ is determined, in a certain energy
interval, by the number of nodes along the (convex) boundary
of the cavity leading to  $K\gg 1$  in the one-channel case
\cite{1}.

For the analysis of transmission and reflection of quantum billiards
with two leads attached to them, it
is convenient to write the $S$ matrix in the following manner \cite{ben}
\beq
\left(
\begin{array}{cc}
S_{m m'} & S_{m n} \\
S_{n m} & S_{n n'}
\end{array}
\right)
\equiv \left(
\begin{array}{cc}
r          & t^{\prime} \\
t          & r^{\prime}
\end{array}
\right) \; .
\eeq
Here,  $m    (n)$ denote the channels in  lead $1$ ($2$).
The matrices $r$ and $\; r'$  describe the reflection
in the lead $1$ and $2$, respectively, while the matrices
$t$ and $t^{\prime}$ describe the
transmission from lead $1$  to lead $2$ and vice versa.
The total transmission and reflection probabilities for the
modes $m$ are 
\beq
\label{tr}
T_{m}=\sum_{n=1}^Z |t_{m n}|^2 \quad 
{\rm and} \quad R_{m}=\sum_{m'=1}^Z
|r_{m m'}|^2
\; ,
\eeq
respectively. As shown by Landauer \cite{ben,alh}, 
the conductance G is proportional 
to the sum of the transmission probabilities,
\beq
\label{cond}
G=  \sum_{m}T_{m} 
\eeq
in the units used by us (see above).
The fluctuations in the transmission and reflection amplitudes 
can be analysed  by means of a Fourier transformation
\beq
\label{ft}
|t_{m n} (L)|^2={\Bigg |}\int dk \; t_{m n} (k) 
e^{-ikL}{\Bigg |}^2
={\Bigg |}\int \frac{dE}{2\sqrt{E}} \; t_{m n} (E)
e^{-i\sqrt{E}L}{\Bigg |}^2 \; .
\eeq
The sum
\beq
\label{tft} 
P(L)=\sum_{m n}|t_{m n} (L)|^2  
\eeq 
is called the  power spectrum  \cite{gutz}.
An analogous expression can be written down for the 
Fourier transform of the reflection amplitudes.

\section{Numerical results}

\subsection{Quantum mechanical and classical calculations}

We study a stadium of Bunimovich type (linear length  
$S=3 \pi /(\pi+1)$ and radius  $R=S$) in the ballistic regime 
with different positions of the
attached leads. The results  are compared with
those of a semicircle $(R=3)$ with an internal scatterer (SIS)
and  leads attached at both ends of the convex boundary.

In the first case (B1) of the Bunimovich billiard, the leads are
attached in the middle of each convex boundary  in the same direction 
so that the WGM are favoured for the conductance, i.e. the coupling
matrix elements $W_R^{c({\rm d})}$ of the WGM with the channel
waves are large. This case is in full analogy to the SIS. In the
second case (B2), the leads are attached in the middle of each
linear boundary in opposite directions
so that the BBM are favoured for the conductance. In
the third case (B3), the leads are attached at the convex boundary
in different directions in such a manner that neither WGM nor BBM
are favoured for the conductance. We compared the results with
those of classical calculations for billiards with the same
geometry.

To find the poles of the $S$ matrix, we use the method of the exterior
complex scaling in combination with the finite element method.
For details see \cite{pese}. The results of the calculations
give us the values $E_R - \frac{i}{2} \Gamma_R = \tilde E_R (E=E_R)
- \frac{i}{2} \tilde \Gamma_R (E=E_R)$ (solutions of the fixed-point
equations, see \cite{ro91,1,ro01}).
The conductance is calculated by direct solving the Schr\"odinger equation 
in a discretized space according to the method suggested in Ref.  \cite{ando}.
The essential ingredients are the conductance formula
(\ref{tr}) and (\ref{cond}), the relation of
transmission coefficients to the $S$ matrix and the corresponding 
Green function, and a recursive calculation of the Green function.

The Fourier analysis of  the transmission and reflection amplitudes
provides us the power spectrum $P(L)$ for
one open channel (one propagating mode, $m=n=1$) and for two open
channels ($m=1,2$, $n=1,2$) in both leads 
according   to Eqs. (\ref{ft}) and (\ref{tft}).

In the classical calculations, we consider the  motion of a
free particle inside  the billiard.
The potential is assumed to be zero inside the billiard and the
boundaries are  mirrors for the motion of the particle along
trajectories which are  calculated from the laws of the geometric optics.
Each trajectory starts at some arbitrarily chosen initial point $(x_0,y_0)$
of the attached leads with an angle  $\Phi_0$  which
characterizes the direction of the motion. We choose $1000\times 1000$
initial conditions to calculate the distribution (histogram) of
the trajectories which contribute to the transport.
The classical conductance is defined as the number 
of trajectories starting at one of the leads and escaping from the other one, 
divided  by  the total number of  trajectories ($10^6$). 
Trajectories with bounces  at the convex boundary only are called 
{\it trajectories of WGM type} in the following. 
The number of such trajectories decreases with increasing number 
of bounces, see e.g. Fig. 3 in Ref. \cite{1}.

\subsection{Eigenvalue pictures}

Fig. \ref{fig:pol} shows the results of numerical calculations
for the four quantum billiards mentioned above. For the SIS we
find, as in \cite{1},  bands $A$, $B$ and $C$ of overlapping resonance 
states whose  widths are large, while the widths of all the other states are
small (Fig. \ref{fig:pol}.a). 
The short-lived states of the bands $A$, $B$ and $C$ start 
at the opening of
thresholds at $E=\pi^2, ~4\pi^2, ~9\pi^2$, respectively. 
At energies $E > 4\pi^2$, we have
channel wave functions which are  effectively coupled to one 
channel mode. 
They exist besides the less coupled channel wave functions  \cite{1}.
In an analogous manner, the channel wave functions  may be
effectively coupled to one or two modes  beyond $E=9\pi ^2$.

The eigenvalue picture Fig. \ref{fig:pol}.a is the
result of resonance trapping occurring according to Eq. (\ref{tra})
and of channel channel coupling, see section 2. The
states with large widths are localized along the convex boundary of the
cavity (Fig. \ref{fig:pol}.b and \cite{1}). They are modes of WGM type.
The states of the band $A$
have a strong overlap with effectively one open channel  in both leads
at all energies. The states of the second band $B$ are 
related to effectively two open channels in each lead 
while the states of the band $C$ are related to  three 
channels. At higher energies, the states of the different bands
interact with one another, 
and the structure of the resonance wave
functions represents a mixture of the states of different bands.

The results for the billiard B1 (Fig. \ref{fig:pol}.c,d) are very
similar to those for the SIS. The difference in the widths of the
short-lived and long-lived states is however smaller and the wave
functions of the B1 are less localized than those of the SIS. 
This is caused by modes of WGM type  localized along 
the lower  boundary of the B1. Such modes are 
coupled weakly  to the attached leads.

The attaching of the leads at the linear boundary (B2;
~Fig. \ref{fig:pol}.e,f) gives rise to  large widths for
states  of the  BBM type.
The differences between the WGM and BBM consist in the following.
\begin{enumerate}
\item[--]
The WGM  are localized along the  boundary of the cavity while
the BBM are localized inside the billiard along the direct connection
between the two attached leads.
\item[--]
The number of the BBM as well as the degree of their overlapping
in the complex plane  are smaller than the corresponding values for the WGM
in the same cavity.
\item[--]
The BBM do trap the other states less than the WGM do, i.e.
some other states (in particular those of WGM type)
still survive in the B2 with small but nonzero widths. These states
take, for example,  altogether about $ 17\%$ of the total sum  
$\Sigma_R \tilde \Gamma_R$ of the widths for $\pi^2<E<4\pi^2$. 
\end{enumerate}

In the B3  billiard (Fig. \ref{fig:pol}.g,h) the  coupling matrix elements
$W_R^{c({\rm d})}$  of  the WGM are large but with different phase
in relation to the two leads. As in the two foregoing cases, the
poles with the largest widths are connected with one another
for illustration. The wave function of one of the states is shown 
in Fig. \ref{fig:pol}.h which is, however, 
less representative for a certain group of
states than in the foregoing cases (Figs. \ref{fig:pol}.b, d, f).

\subsection{Power spectra and classical trajectories}

In Fig. \ref{fig:con}, we present the (energy dependent) conductance $G$
calculated quantum mechanically and the mean value $\bar G$ of the
conductance. Furthermore we show, in this figure,  
the corresponding  power spectra $P(L)$ and
the histograms of trajectories  calculated
classically for transmission as a function of the length $L$
of the path for the four different types of billiards.
The results display a remarkable and surprisingly good
agreement between the quantum mechanical results
of the Fourier analysis and the classical results in spite  of the small value
of the wave vector $k$ of the propagating waves.

In the SIS and B1 with dominant WGM, the
largest peak in the $P(L)$ spectrum can be identified
with the length of  the path of the WGM trajectories calculated
classically.   In contrast to the SIS, the classical trajectories 
of the B1 with small $L$ are split into 
two parts corresponding to one bounce at the convex boundary and
to two bounces, respectively. 
The number of paths with two bounces is much smaller than that with one bounce
in full agreement with the expectation.
Typical pictures of these  trajectories are shown 
near to the corresponding peaks in the histogram  Fig. \ref{fig:con}.f.
In both cases, SIS and B1, smaller peaks can be identified 
with  other trajectories which are
however of minor importance for the transport.
The energy dependent conductivity $G$ 
(especially of the SIS,  Fig. \ref{fig:con}.a) reflects the strong
channel channel coupling between the two channel modes at $E > 4 \pi ^2$
which is responsible for the high conductance also at higher energies.

In the B2 billiard, two peaks of comparable
height appear in the $P(L)$ spectrum  (Fig. \ref{fig:con}.h).
A representative wave function of the states belonging to the first peak
is displayed in Fig. \ref{fig:pol}.f while another one for the second peak
is shown in Fig. \ref{fig:dbb}. In the first case, channel channel coupling
creates effectively one channel while there are effectively two channels
in the second case.
The corresponding lengths  $L$  differ by about a factor 2. This is in
agreement with the differences of the paths calculated classically for the two
highest peaks in   Fig. \ref{fig:con}.i without any bouncing and with two
bouncings at the convex boundary of the cavity, respectively.
The conductivity of the B2 billiard (Fig. \ref{fig:con}.g)
is  determined only partly by channel channel coupling.

The differences of the BBM case (B2)
to the two WGM cases (SIS and B1) consist in the following.
\begin{enumerate}
\item[--]
The $P(L)$ spectrum is dominated by one peak at small $L$ 
in the WGM cases, while there are two peaks
of less height in the BBM case.

\item[--]
The $|t_{11}(L)|^2$ spectra (defined in the energy range $\pi^2 < E < 4 \pi^2$)
are dominated in all three cases by  one peak
at small $L$ the height of which is, however, smaller in the BBM case than
in the WGM cases.

\item[--]
$\bar G(1)$ and  $\bar G(2)$ are smaller  in the BBM case
than in the two WGM cases.

\end{enumerate}

The results for the B3 billiard do not show any pronounced peaks
in the power spectrum at short lengths $L$. The mean conductivity
is near to the classical value in accordance with 
the prediction of random matrix theory \cite{ben}.  

In Fig.  \ref{fig:ref}, we present the  power spectra of the
reflection amplitudes for the four billiards
studied above. Additionally, we show in each case the wave function
of a state lying at the energy   where the conductance is minimal.
In contrast to the power spectra $P(L)$ of the conductance, the power spectra
of the reflection show more pronounced  peaks for the B2 and B3 billiards
than for the SIS and B1. They appear at comparably large $L$.
In any case, the peaks
in the power spectra of the conductance and reflection
are at different lengths $L$ for every cavity. This holds especially for the
first peak of the power spectrum for the  reflection   
in the B2 which lies between the two BBM peaks of the power spectrum 
for the transmission.

In Table \ref{tab}, the results obtained
for the conductivity from the quantum mechanical 
calculations are compared with those from the  classical calculations.  
It is remarkable that the conductivity is determined,
to a great deal, by trajectories
of WGM type in the classical calculations as well.
Their contribution is about 45 \% and 28 \%
of trajectories for the SIS and the B1, respectively. 
It is smaller in the latter case since the boundary of the B1 is not convex
everywhere in contrast to that of the SIS. In the quantum mechanical
calculations for the SIS and B1, the conductivity is maximum at 
low energy  with one open
channel. It decreases with increasing energy.

The small conductivity of the B2 at low energy (Table \ref{tab}, 
one channel) is rather unexpected
at first sight, since  the classical path corresponding to the 
BBM trajectories is the direct one. Their contribution   is, in the
classical calculations,
however only about 7 \% of the total number of 
the transmitted trajectories, whereas the WGM trajectories
contribute about 11 \%. That means the trajectories  occupy,
to a large part, the available inner space of the billiard, 
resulting in a reduction of  the conductivity. This tendency can be seen 
also in the quantum mechanical  calculations,
see  Figs. \ref{fig:pol}.e and \ref{fig:dbb}.

Further, we calculated quantum mechanically the
$\overline{|t_{nm}|^2}$  for five open channels in each lead 
($n,m=$ 1,...,5) in the energy range $25 \pi^2 < E < 36 \pi^2$ 
for the B1 billiard (Table \ref{tab2}),
the corresponding Fourier transforms of the $|t_{nm}|^2$ 
(Fig. \ref{fig:trans}) and, for illustration,  
the number of classical trajectories  travelling through the billiard
(Fig. \ref{fig:clas}). In the classical calculations, we
included only trajectories with lengths smaller than 20
according to the results shown in  Fig. \ref{fig:con}.f.  The angle
$\Phi$ is determined by the
trajectory going into the billiard ($\Phi_{\rm in}$) or leaving it
($\Phi_{\rm out}$). Using the quantum mechanical correspondence between energy
and angle $\Phi=arctan \Big(\pi n \Big/ \sqrt{E-(\pi n)^2}\Big)$,
we divide the $|\Phi_{\rm in}| - |\Phi_{\rm out}|$ plane into $5 \times 5$
blocks corresponding to the  transmissions  $|t_{nm}|^2$. 
The angle   $\Phi$ is measured with respect to
the normal of the attachment line between lead and billiard.
The trajectories which enter and leave the cavity at an  angle
around $\Phi \approx 0$ can be identified with trajectories of WGM type.
The dark straight line can be associated with trajectories which bounce once
the linear boundary of the billiard ($\Phi \approx \pi / 4$). 
Most trajectories with large angles are longer than 20 
and do not appear  in Fig. \ref{fig:clas} since they are not taken into
account in the classical calculations.

\section {Discussion of the results}

Comparing the results for the different billiards (performed for the 
ballistic regime), we see the strong influence
of the lead orientation onto the resonance wave functions and the conductance
or reflection. The results can be understood on the basis of Eq.
(\ref{smat}) which involves the
coupling coefficients $\tilde W_R^c$ between the {\it resonance} states  
and the channel wave functions in the leads. It follows: 
\begin{enumerate}
\item
The most effective attachment of the leads for a selection of special modes
and a high conductance 
is the symmetrical one with $\tilde W_R^c \approx \tilde W_{R}^{c'}$.

\item
The destructive interferences in the transmission amplitudes  are 
reduced when the  number of states {\it and} channels
is effectively  reduced.

\end{enumerate}

The first  condition is fulfilled for the SIS and the B1 
with selection of the WGM
as well as for the B2 with selection of the BBM.
It is {\it not} fulfilled for the B3
where $\tilde W_R^c$ is large for the WGM along the upper  boundary  but
$\tilde W_R^{c'}$ is small, and vice versa for the WGM along the lower
boundary. 
Although the number of WGM is more or less the same in the B3 as in
the B1, the conductance  is  very different in the two cases.

The second condition is fulfilled to the maximum by resonance trapping.
The differences
in the coupling coefficients  $\tilde W_R^c$ between the {\it resonance} 
states $R$ and the wave functions of the channels $c$ 
are larger than those in the original coupling coefficients 
$ W_R^{c {\rm (d)}}$ between the discrete states and the channels.
A few of the  $\tilde W_R^c$ may reach the maximum possible value determined
by Eq. (\ref{gasu}) while those of the remaining ones approach zero,
meaning that they are almost decoupled from the channels.
Due to the large coupling coefficients between the special states and 
the channel wave functions, 
the channels  are coupled via these  states. As a consequence, not only
the number of states is effectively reduced, {\it but also} the 
number of channels is effectively scaled down. 
In this manner,  a few special 
quantum mechanical states may be selected by the attachment of the leads 
to the cavity whose number is, in any case, smaller than the total number of
states. Further, the special states are coupled mainly  to 
some channels whose number is effectively
smaller than (or  at most equal to)  the total number of open channels
(for illustration see Fig. {\ref{fig:pol}   and \cite{1} for quantum billiards
and also \cite{drokplro} for nuclei). 
Thus, the interferences between the transmission
amplitudes are reduced by the phenomenon of resonance trapping.

Another illustration for the effective reduction of the  
number of channels, to which the special states are coupled,
is shown in Fig. \ref{fig:clas} where the
quantum mechanical transmission matrix elements,
calculated with account of 5 channels (modes) in each lead, are mapped
onto the classical transmission matrix, calculated with account of   
paths shorter than 20. The classical transmission through 
short paths ($L\leq 20$) corresponds to the quantum mechanical 
transmission through 
the special states with  at most four (out of five) channels.  
In the energy region between $25 \pi^2$ and $36 \pi^2$ there are,
in the quantum mechanical calculations with 5 channels,
however also contributions from other states with longer paths to
the transmission (Fig. \ref{fig:trans}).  
While the Fourier transforms of $|t_{mn}|^2$ with $mn =$ 11, 12, 14,
33, 34 and 44  have a well expressed peak around $L\approx 14$ to 15, 
this is not so in the other cases. 
The Fourier transforms with $mn =$ 22, 23 are strongly peaked 
around $L=30$ while those 
with   $mn =$ 13, 15, 24, 25, 35 and 45  are distributed over different
$L>15$ and that with  $mn =$ 55 even over $L>27$. 
As can be seen from these numbers, the quantum mechanical contributions 
with $L<20$ to the conductance are restricted to 4 channels in each lead,
indeed. That is in full accordance to the classical picture.
The increasing contributions to the conductance from states with 
larger $L$ weaken however the channel channel coupling, and 
the effective number of channels approaches the number $Z$ of independent
channels. The results of classical calculations without the 
restriction to small $L$ 
(not shown) correspond to this result of the quantum mechanical
studies.  

According to the numerical results for 
$\overline{|t_{nm}|^2}$ with 5 channels in each lead (Tab. \ref{tab2}), 
the contributions to the conductance 
from the  $|t_{mn}|^2$ with a single peak around  
$L\approx 14$ to 15 are mostly larger than those from  the other $|t_{mn}|^2$.
Nevertheless, the contributions from  states 
with  paths $L>20$ have to be taken into account. 
In all the cavities considered by us, the conductance approaches the 
classical value with increasing 
number of channels (Tab.  \ref{tab}). For the B1 
with 5 channels, we obtain $G/Z= 0.66$. 

According to Eqs.  (\ref{gasu}) and (\ref{gasub}),
the coupling strength between cavity and lead is finite so that the widths
of the special states can reach, by resonance trapping, 
a maximum possible value only. By this, the conductivity is restricted
in value also. In some cases (WGM only along the upper boundary as 
in the SIS), 
the conductivity is enhanced, indeed,  almost up to the maximum  
possible value whereas this is not so in the other cases.
Neither the BBM modes in the B2 nor the WGM modes in the B1 are able to
trap completely the remaining states  which include 
special states of WGM type (along the
lower boundary in the B1 and along the whole  boundary in the B2). 
The maximum value 
of the conductance can therefore not be reached in these cases.
While the special states determine
the average properties  of observables such as  conductance and reflection,  
the trapped states  are responsible for the fluctuations around the mean values.
This result is independent of the existence of an internal scatterer inside
the cavity. More important than the internal scatterer is the convex lower
boundary of the B1 in contrast to the linear lower boundary of the SIS.

Characteristic of  special states of a certain type is  the ratio
$M^{\rm spec} / M$ (where  $M^{\rm spec}$ is the number of
special states and $M$ is the total number of states in a certain
energy interval) as well as the dependence of the coupling matrix
elements $W_R^{c {\rm (d)}}$ on the parameter varied. In the
cases considered in the present paper, not only 
the number of WGM is larger than that of BBM, 
but the WGM overlap  stronger and are more
stable against small shifts of the leads than the BBM (the last
point is studied in \cite{1} for the SIS). While the WGM are able
to trap almost all other states under favourable for them
conditions, the BBM do never  trap the WGM completely (compare
Fig. \ref{fig:pol}.c with \ref{fig:pol}.e and see \cite{1} for the
SIS). These differences are  related to the fact that the
WGM are more strongly localized than the BBM. While the
WGM are localized  along the (convex)  boundary of the system,
the BBM are localized inside the system near to the shortest connection
between the two leads. Deviations from the shortest
distance  appear under the influence of the area  of the billiard. 

In all cases considered by us, the special states  (WGM and  BBM)  
accumulate, by resonance trapping, the major part of the coupling strength 
between system and lead (sum of the  widths of {\it all} states). 
The close correspondence
between the quantum mechanical and classical calculations
is related, at least partly, to the existence of these special states.
Fig. \ref{fig:con} shows the correspondence in relation to the lengths
$L$. Let us now consider the correspondence in relation to the 
lifetimes (widths).

To this aim, we focus  on the B1 and the B2 billiards
in the energy interval between the first ($\pi^2$) and second 
($4\pi^2$) thresholds where the WGM and BBM  states are well
separated from the other resonance states.
In the B1 billiard, the special states consist of eleven WGM.
The average width of these eleven states is
${\bar \Gamma}_{\rm WGM}\approx 6.5$.  Their contribution
to the total coupling strength between system 
and environment, $\sum_R \tilde \Gamma_R=76.6$, is $93\%$. 
In the B2 billiard, five special states of  BBM type  accumulate
$83\%$ of the total coupling strength. Here,
${\bar \Gamma}_{\rm BBM}\approx 12.6$ and 
$\sum_R \tilde \Gamma_R=76.1$.

To get an estimation for the mean width   
$\langle \Gamma \rangle $ of the resonance states 
in a quantum billiard ({\it without}
taking into account the mixing of the resonance states via the continuum) 
we use the expression \cite{gutz}
\beq
\rho= M/\Delta E = \frac{A}{4\pi}\frac{2m_{\rm eff}}{\hbar^2} = 
\frac{A}{4\pi}
\eeq
for the level density
(in units $\hbar^2/2m_{\rm eff} =1$, see Section \ref{basic}).
Here, $\Delta E$ is the energy interval considered, $M$ is the
number of states and $A$ is the area of the quantum billiard. 
The total number of resonance states  between the
first and second threshold $(\pi^2$ and $4\pi^2$)
is $M=3\pi^2 \, A /(4\pi)  \approx 67$ for both billiards, since they have 
the same area.  
Also the average coupling strength is approximately the same 
for the two billiards, see Eqs. (\ref{gasu}) and (\ref{gasub}). 
The estimation yields 
\beq
\langle \Gamma \rangle= \sum_R \Gamma^{\rm d}_R/M \approx
\sum_R \tilde \Gamma_R/M \approx 1.1 \; . 
\label{random}
\eeq 
It is interesting to compare the quantum mechanical values 
\cite{remark}
\beq
\langle \Gamma \rangle =  {1\over {\tau}} \quad  \quad \quad 
\bar\Gamma_{\rm S} = {1\over{\tau_{\rm S}}}
\label{tau}
\eeq
for the mean lifetimes with those obtained from the classical calculations
for the  time of flight
where S stands for WGM and BBM, respectively.  
A rough estimation of the flight time  for a particle along the WGM or BBM
trajectories gives 
$\tau^{\rm cl} = L^{\rm cl}/v = L^{\rm cl}/k_n = 
L^{\rm cl}/\sqrt{E-n^2\pi^2}$ and therefore
\beq
{\langle \Gamma^{\rm cl} \rangle} = {{\sqrt{E-n^2\pi^2}}\over {L^{\rm cl}}} 
\; .
\label{estim}
\eeq
We get   ${\langle \Gamma^{\rm cl}_{\rm WGM} \rangle}
\approx 0.5 $ for the WGM trajectories with $L^{\rm cl}_{\rm WGM}=3\pi+2$ 
and ${\langle \Gamma^{\rm cl}_{\rm BBM} \rangle}
\approx 0.8 $ for the BBM trajectories with $L^{\rm cl}_{\rm BBM}=
6\pi/(\pi+1)+2$  and maximum energy.
These values  are of the same order of magnitude as the 
$\langle \Gamma \rangle$ calculated quantum mechanically. 
The values $\bar\Gamma_{\rm S}$ of the special states, however,  
are much larger due to resonance trapping. It is
$\bar\Gamma_{\rm BBM} / \bar\Gamma_{\rm WGM} 
\approx \langle \Gamma_{\rm BBM}^{\rm cl} \rangle / 
\langle \Gamma_{\rm WGM}^{\rm cl} \rangle = 
L^{\rm cl}_{\rm WGM} / L^{\rm cl}_{\rm BBM} $ .
The relation  
\beq
\bar \Gamma_{\rm S} \; \propto \; 
{1 \over L^{\rm cl}_{\rm S}}
\; \propto \; \langle \Gamma^{\rm cl}_{\rm S} \rangle
\label{time}
\eeq
holds in all our calculations, see e.g. Fig. 2.e in Ref. \cite{1}, while 
$\langle \Gamma \rangle$, Eq. (\ref{random}), is related to the area 
of the cavity and is (almost)
independent of the manner the leads are attached to it. That means, 
$\langle \Gamma \rangle$ is {\it not} related to any special $L$
in contrast to $\bar \Gamma_{\rm S}$.
The shortened lifetimes $\tau_{\rm S}$ are an  expression for  the
collective properties of the special states which  result from the 
quantum mechanical phenomenon of resonance trapping.
They allow, under certain
conditions, an enhancement of the conductance, as discussed above.

All the results obtained in the present study show the close
correspondence between the classical  and the quantum mechanical
characteristics for the transport through billiards of different shape. 
This correspondence is achieved by the dynamics of open quantum systems
which is determined by the shape of the cavity and
the position of the attachment of the leads to it. The dynamics 
can be understood on the basis of the resonance reaction part  
(\ref{smat}) of the $S$ matrix that involves not only the 
wave functions of the states
of the closed system but also the influence of the environment 
onto the properties of the system including the phenomenon of resonance
trapping following from Eqs. (\ref{gasu}) and (\ref{gasub}).

\section{Conclusions}

For the Bunimovich stadium with two attached leads
we  have  calculated  energies,  wave functions and coupling coefficients
to the environment (widths). As a result, all these values may change strongly
by varying the position of the attached leads.
The changes can be seen in observables such as  conductance
or reflection.

Our study  shows that special states exist in  open quantum billiards.
These states have individual (non-generic) properties characteristic of both
the geometry of the system and the position of the attached leads.
They  have large widths (small lifetimes) due to
trapping other states most of which have lost  their
individual properties they had in the closed cavity, see e.g. \cite{pese}. 
The wave functions of the
special states are localized  while those of the 
trapped states are distributed over the whole cavity.
The special states determine, as a rule, the mean properties of observables 
(such as the conductance) while
the trapped states are responsible for the fluctuations around the mean
values. The contribution of  special states to physically relevant
values can be enhanced  by the attachment of  leads to the
billiard in such a manner that the coupling of these states to the channel
wave functions is favoured.
These results are in qualitative agreement with experimental data obtained
from quantum dots with different lead alignments \cite{a2}.
Examples of special states are, above all,  the WGM  studied in this paper.
The BBM are less stable.

The most interesting result of our study is the relation between classical 
and quantum mechanical properties of the open microwave cavities.
The short-lived special states are localized 
around the  classical paths with very few bounces at the boundary
and are coupled strongly to a small number of effective channels.
The lifetimes of these states  depend  on the geometry of
the billiard: they are proportional to the lengths of the
classical trajectories. In contrast to this, the long-lived trapped states
are delocalized (i.e. distributed over more or less the whole area of the 
billiard) and coupled very weakly to  all channels.
It should be underlined  that the
coherent short-lived and incoherent long-lived resonance states
exist always together at the same energy. 
Only the long-lived trapped states can cause the randomness of the system.

We conclude the following.
The classical properties of dynamical systems are manifest in quantum
mechanical characteristics of open systems even at low 
energy where the level density and the number of open channels are small. 
The classical properties are related, above all, to some special states
that exist in the closed system and whose special features may be 
strengthened by coupling the system to an environment
by an appropriate position of the leads.
This enhancement is caused by the phenomenon of resonance trapping.
It is the stronger the larger the number of states as well as the number of 
open channels is. It  is accompanied (i) by
the formation of long-lived states in the same energy region which contribute
incoherently to the observable values
and (ii) by a reducing of the effective number of channels
for the decay of the special states. Due to the destructive interferences 
between  the short-lived special states and the long-lived  trapped states,
an enhancement (reduction) of observable values appears only at low level 
density. This result discussed in the present paper on the example of the
transmission (reflection) through quantum billiards,  
is expected to be true also for other observables   
and, above all, for real quantum systems such as quantum dots.

\vspace{.5cm}

{\it Acknowledgement:}
We thank D. V. Savin and V.V. Sokolov for useful discussions
and H. Schomerus for the critical reading of the manuscript. 
This work was supported in part by the Russian Foundation for Basic
Research Grant No.  01-02-16077, the Heisenberg-Landau program
of the BLTP, JINR, the Czech grant GAAV A1048101 and
by the "Foundation for Theoretical Physics" in Slemeno, Czech Republic.

\vspace{.5cm}

{\it E-mail addresses:}\\
rashid@thsun1.jinr.ru\\
knp@tnp.krasn.ru \\
rotter@mpipks-dresden.mpg.de\\
seba@fzu.cz \\

\newpage

\begin{figure}
\caption{
The poles of the $S$ matrix and a representative picture
$|\Phi_R|^2$ of the wave functions of the
short-lived states (belonging to the group A)
for the SIS (a, b), B1 (c, d), B2 (e, f) and B3 (g, h).
The poles of the $S$ matrix (denoted by stars) far from the real axis
are  connected by lines for guiding the eyes.
The energies and widths are in units of the width of the attached waveguide.
}
\label{fig:pol}
\end{figure}

\begin{figure}
\caption{
The conductance
 $G(E) = \sum_{m,n}|t_{mn}|^2  $ (calculated quantum mechanically),
the corresponding power spectrum $P(L)$, and
the histogram of the (classically calculated) trajectories
for conductance  as a function of the length $L$
for the SIS (a, b, c), B1 (d, e, f), B2 (g, h, i), and B3 (j, k, l).
In (a, d, g, j),
$\overline G(1)$ and $\overline G(2)$ denote the mean
value of the conductance
in the energy intervals $\pi^2 < E < 4 \pi^2$ and $4 \pi^2 < E < 9 \pi^2$,
respectively. In (b, e, h, k), the total power 
spectrum $P_{tot}(L)=\sum_{m,n}|t_{mn}(L)|^2$
of the transmission amplitudes (thick lines)
in the energy interval $\pi^2 < E < 9 \pi^2$ 
and the power spectrum of the transmission amplitudes
$|t_{11}(L)|^2$ in the energy region $ \pi^2 < E < 4 \pi^2$ with two open
channels in each lead (dash-dotted lines) are shown.
Typical classical trajectories are displayed
near to the corresponding bins in (c,  f, i, l).
Note the different scales of $P(L)$ in (b, e, h, k).
}
\label{fig:con}
\end{figure}

\begin{figure}
\caption{ A representative picture $|\Phi_R|^2$ for the wave functions of the
states which belong to the second peak of $P(L)$ at $L\approx 16$ for the B2.}
\label{fig:dbb}
\end{figure}

\begin{figure}
\caption{The power spectrum $P_{tot}(L)=\sum_{m,n}|r_{mn}(L)|^2$
for the reflection amplitudes (thick lines)
in the energy region $\pi^2 < E < 9\pi^2$
and $|r_{11}(L)|^2$ in the energy region $\pi^2 < E < 4\pi^2$ with two open
channels in each lead (dash-dotted lines) for the SIS (a), B1 (c),
B2 (e) and B3 (g). The wave function $|\Phi_R|^2$ of a state, lying at an
energy where the conductance is small,
for the SIS (b), B1 (d), B2 (f) and B3 (h).
}
\label{fig:ref}
\end{figure}

\begin{figure}
\caption{
The power spectra $p(L)\equiv |t_{nm}(L)|^2$ for the B1 billiard  
in the energy region $25 \pi^2 < E < 36 \pi^2$ with five open
channels in each lead. In the figure, only those power spectra are shown for
which the height of at least one peak is larger than 0.5.
}
\label{fig:trans}
\end{figure}

\begin{figure}
\caption{
The  transition matrix calculated classically for the B1  as a function 
of the angle of the
ingoing and outgoing waves at which the classical trajectories
pass  the attachment of the leads. The length of the trajectories is
restricted to  $L \leq 20$.  The 
transmission coefficients $t_{nm}$ ($n,m=$ 1,..., 5) of the
quantum mechanical calculations  for 5 modes in each lead  (Table \ref{tab2}) 
can be mapped  onto the figure  as indicated. 
}
\label{fig:clas}
\end{figure}

\begin{table}
\caption{ 
The conductance $G/Z$ for different billiards with different 
number $Z$ of channels} 
\vspace*{.5cm}
 \label{tab}
 \begin{tabular}{ccccc}
\hspace*{2cm} billiard type & 1 channel   &  2 channels & 3 channels &  
classical \hspace*{2cm}
\\
\hline
 & & &  \\
\hspace*{2cm} SIS & 0.87 & 0.75    & 0.74 & 0.66 \hspace*{2cm}
\\
\hspace*{2cm} B1 & 0.74 & 0.73  & 0.65 &  0.63 \hspace*{2cm}
\\
\hspace*{2cm} B2  &  0.46 & 0.56  & 0.56  &  0.57 \hspace*{2cm}
 \\
\hspace*{2cm}  B3 & 0.49 & 0.46  & 0.56  & 0.53 \hspace*{2cm}
\\
& & &  \\
 \end{tabular}
\end{table}

\vspace{1cm}

\begin{table}
\caption{ 
The values  $\overline{|t_{nm}|^2}$  for the B1 billiard with 
$n,m=$ 1,...,5  }
\vspace*{.5cm}
 \label{tab2}
 \begin{tabular}{ccc}
\hspace*{6cm} $n$    &  $m$ & $\overline{|t_{nm}|^2}$ \hspace*{6cm}
\\
\hline
 & &  \\
\hspace*{6cm} 1 & 1 & .45    \hspace*{6cm}
\\
\hspace*{6cm} 1 & 2 &  .10 \hspace*{6cm}
\\
\hspace*{6cm}  1 &  3 & .05 \hspace*{6cm}
 \\
\hspace*{6cm}  1 & 4 &  .06 \hspace*{6cm}
\\
\hspace*{6cm}  1 &  5 &  .10 \hspace*{6cm}
 \\
& &   \\
\hspace*{6cm} 2 & 2 & .18 \hspace*{6cm}
\\
\hspace*{6cm}  2 &  3 & .09 \hspace*{6cm}
 \\
\hspace*{6cm}  2 & 4 &  .17 \hspace*{6cm}
\\
\hspace*{6cm}  2 &  5 & .10 \hspace*{6cm}
 \\
& &   \\
\hspace*{6cm} 3  &  3 &  .31 \hspace*{6cm}
 \\
\hspace*{6cm}  3 & 4 & .14 \hspace*{6cm}
\\
\hspace*{6cm}  3 &  5 & .08 \hspace*{6cm}
 \\
& &   \\
\hspace*{6cm} 4  & 4 &  .31 \hspace*{6cm}
\\
\hspace*{6cm}  4 &  5 & .07 \hspace*{6cm}
 \\
& &   \\
\hspace*{6cm} 5  &  5 & .16 \hspace*{6cm}
 \\
& &   \\
 \end{tabular}
\end{table}


\begin{thebibliography}{99}
\bibitem{remark} The relation (\ref{tau}) is true in atomic units 
$\hbar = |e| = m =1$ which may differ from the units 
$\hbar^2/2m_{\rm eff} =1$ used in the present paper. The result discussed by
us is, however, not influenced by this difference due to
the following fact: $\langle \Gamma \rangle $ 
is (more or less) the same for the different
billiards considered in the present paper,
while the corresponding $\langle \Gamma^{\rm cl} \rangle$  differ by a
factor of almost 2.
\end{thebibliography}
\end{document}